\documentclass[12pt]{iopart}
\usepackage{hyperref}
\usepackage{epsfig}
\usepackage[numbers]{natbib}

\usepackage{setstack}
\usepackage{iopams}
\usepackage[latin1]{inputenc}
\usepackage{url}

 \providecommand{\email}[1]{\href{mailto:#1}{\texttt{#1}}}
\providecommand{\dprod}{\! \cdot \!}%
\providecommand{\wprod}{\! \wedge \!}
\begin{document}
%

\title{Maxwell's equations in 4-dimensional Euclidean space}
\author{Jos\'e B. Almeida}
\address{Universidade do Minho, Departamento de F\'isica,
4710-057 Braga, Portugal.}

\ead{\email{bda@fisica.uminho.pt}}


\date{\today}

\begin{abstract}                
The paper formulates Maxwell's equations in 4-dimensional Euclidean space by
embedding the electromagnetic vector potential in the frame vector $g_0$.
Relativistic electrodynamics is the first problem tackled; in spite of using a
geometry radically different from that of special relativity, the paper derives
relativistic electrodynamics from space curvature. Maxwell's equations are then
formulated and solved for free space providing solutions which rotate the
vector potential on a plane; these solutions are shown equivalent to the usual
spacetime formulation and are then discussed in terms of the hypersphere model
of the Universe recently proposed by the author.
\end{abstract}
\pacs{41.20.Jb, 02.40.Dr, 04.20.Cv}

%
\maketitle

\section{Introduction}
In recent years I've been proposing an extension of optics to 4-dimensional
space, naturally calling this discipline \emph{4-dimensional optics} (4DO).
Initially this was proposed as an alternative formulation for relativistic
problems; two papers that use this view and provide a good introduction are
\citet{Almeida02:2, Almeida01}. The latter of these references has some flaws
in its approach to electromagnetism which will be corrected here; it also
suggested a metric for vacuum which was superseded by a more appropriate one in
\cite{Almeida04:1}. In this very recent work I made some cosmological
predictions arising from a development of 4DO, which had so far been impossible
with a general relativity approach, showing that dark matter is really
unnecessary for the explanation of observations if one accepts 4-space as being
Euclidean with coordinate $x^0$ the radius of an hypersphere. The latter work
left unexplained the fact that photons were constrained to great circles in a
4-dimensional hypersphere rather than following straight line geodesics.

The explanation of photon behaviour calls for a full exposition of Maxwell's
equations in 4DO context, which has not yet been done in a formal way. The
present paper is a presentation of electromagnetism in 4DO space, introducing
the electromagnetic vector potential as part of the space frame and deriving
Maxwell's equations in a natural way. The solution of Maxwell's equations
leading to electromagnetic waves is then shown to be bound to great circles on
the 4D hypersphere, thus solving the difficulty in \citet{Almeida04:1}.

In the exposition I will make full use of an extraordinary and little known
mathematical tool called geometric (Clifford) algebra, which received an
important thrust with the works of David Hestenes \cite{Hestenes84}. A good
introduction to geometric algebra can be found in \citet{Gull93} and in the
following paragraphs I will use the notation and conventions of the latter.
Expressing Maxwell's equations in the formalism of geometric algebra is not
new; the Cambridge Group responsible for the reference above uses this approach
in one of their courses \cite{Lasenby99} for the relativistic formulation of
those equations, which are then condensed in the extraordinarily compact
equation
\begin{equation}
    \nabla^2 A = J.
\end{equation}
Although this formulation is valid for Minkowski spacetime, with a different
signature from 4DO, I shall follow closely that group's exposition.

Before embarking into the transposition of electromagnetism to 4DO the paper
makes a brief introduction to geometric algebra and makes a quick revision of
4DO's principles, using the opportunity to express them in the former's
formalism.
\section{Introduction to geometric algebra}
The geometric algebra of Euclidean 4-space $\mathcal{G}_4$ is generated by the
frame of orthonormal vectors $\{\sigma_\mu \}$, $\mu = 0 \ldots 3$, verifying
the relation
\begin{equation}
    \sigma_\mu \dprod \sigma_\nu = \frac{1}{2}\, (\sigma_\mu
    \sigma_\nu + \sigma_\nu \sigma_\mu ) = \delta_{\mu \nu}.
\end{equation}
The algebra is 16-dimensional and is spanned by the basis
\begin{equation}
    \begin{array}{ccccc}
    1, & \{\sigma_\mu \}, & \{\sigma_\mu \sigma_\nu \}, & \{
    \sigma_\mu I \}, & I, \\
    \mathrm{1~ scalar} & \mathrm{4~ vectors} & \mathrm{6~ bivectors} &
    \mathrm{4~ trivectors} & \mathrm{1~ fourvector}
    \end{array}
\end{equation}
where $I \equiv \sigma_0 \sigma_1 \sigma_2 \sigma_3$ is also called the
pseudoscalar. The elements of this basis are such that all vectors and the
pseudoscalar square to unity
\begin{equation}
 (\sigma_\mu)^2 = 1,~~~~ I^2 =1;
\end{equation}
and all bivectors and trivectors square to $-1$
\begin{equation}
    (\sigma_\mu \sigma_\nu)^2 = -1,~~~~
    (\sigma_\mu I)^2 = -1.
\end{equation}
It will be convenient to shorten the product of basis vectors with a
multi-index compact notation; for instance $\sigma_\mu \sigma_\nu \equiv
\sigma_{\mu \nu}$.

The geometric product of any two vectors $a = a^\mu \sigma_\mu$ and $b = b^\nu
\sigma_\nu$ can be decomposed into a symmetric part, a scalar called the inner
product, and an anti-symmetric part, a bivector called the exterior product.
\begin{equation}
    ab = a \dprod b + a \wprod b,~~~~ ba = a \dprod b - a \wprod b.
\end{equation}
Reversing the definition one can write internal and exterior products as
\begin{equation}
    a \dprod b = \frac{1}{2}\, (ab + ba),~~~~ a \wprod b = \frac{1}{2}\, (ab -
    ba).
\end{equation}

The exponential of bivectors is especially important and deserves an
explanation here. If $u$ is a bivector or trivector such that $u^2 = -1$ and
$\theta$ is a scalar
\begin{eqnarray}
   \mathrm{e}^{u \theta} &= &1 + u \theta -\frac{\theta^2}{2!} - u
    \frac{\theta^3}{3!} + \frac{\theta^4}{4!} + \ldots \nonumber \\
    &= &1 - \frac{\theta^2}{2!} +\frac{\theta^4}{4!}- \ldots \{=
    \cos \theta \} \\
    &&+ u \theta - u \frac{\theta^3}{3!} + \ldots \{= u \sin
    \theta\}\nonumber \\
    &= & \cos \theta + u \sin \theta. \nonumber
\end{eqnarray}
Although less important in the present work, the exponential of vectors and
fourvector can also be defined; if $h$ is a vector or fourvector such that $h^2
=1$
\begin{eqnarray}
    \mathrm{e}^{h \theta} &= &1 + h \theta +\frac{\theta^2}{2!} + h
    \frac{\theta^3}{3!} + \frac{\theta^4}{4!} + \ldots \nonumber \\
    &= &1 + \frac{\theta^2}{2!} +\frac{\theta^4}{4!}+ \ldots \{=
    \cosh \theta \}\\
    &&+ h \theta + h \frac{\theta^3}{3!} + \ldots \{= h \sinh
    \theta\} \nonumber \\
    &= & \cosh \theta + h \sinh \theta. \nonumber
\end{eqnarray}
The exponential of bivectors is useful for defining rotations; a rotation of
vector $a$ by angle $\theta$ on the $\sigma_{12}$ plane is performed by
\begin{equation}
    a' = \mathrm{e}^{-\sigma_{12} \theta/2} a
    \mathrm{e}^{\sigma_{12} \theta/2}= R a \tilde{R};
\end{equation}
the tilde denotes reversion, reversing the order of all products. As a check we
make $a = \sigma_1$
\begin{eqnarray}
    \mathrm{e}^{-\sigma_{12} \theta/2} \sigma_1
    \mathrm{e}^{\sigma_{12} \theta/2} &=&
    \left(\cos \frac{\theta}{2} - \sigma_{12}
    \sin \frac{\theta}{2}\right) \sigma_1
    \left(\cos \frac{\theta}{2} + \sigma_{12} \sin
    \frac{\theta}{2}\right)\nonumber \\
    &=& \cos \theta \sigma_1 + \sin \theta \sigma_2.
\end{eqnarray}
Similarly, if we had made $a = \sigma_2$, the result would have been $-\sin
\theta \sigma_1 + \cos \theta \sigma_2$.

If we use $B$ to represent a bivector and define its norm by $|B| = (B
\tilde{B})^{1/2}$, a general rotation in 4-space is represented by the rotor
\begin{equation}
    R \equiv e^{-B/2} = \cos(|B|/2) -  \frac{B}{|B|}
    \sin(|B|/2).
\end{equation}
The rotation angle is $|B|$ and it is performed on the plane defined by $B$. A
rotor is defined as a unitary even multivector (a multivector with even grade
components only); we are particularly interested in rotors with scalar and
bivector components. It is more general to define a rotation by a plane
(bivector) then by an axis (vector) because the latter only works in 3D while
the former is applicable in any dimension.

In a general situation the frame may not be orthonormed and we will generally
define this frame by expressing its vectors in the fiducial frame $\sigma_\mu$
\cite{Hestenes86:2}
\begin{equation}
    g_\mu = {h^\alpha}_\mu \sigma_\alpha;
\end{equation}
where ${h^\alpha}_\mu$ is called the fiducial tensor. The metric tensor is then
defined by the inner products of the frame vectors
\begin{equation}
    \label{eq:metrictensor}
    g_{\mu \nu} = g_\mu \dprod g_\nu.
\end{equation}
Complementary we can then define the reciprocal frame by the relation
\begin{equation}
    g^\mu \dprod g_\nu = {\delta^\mu}_\nu.
\end{equation}
This method for frame and metric definition is absolutely general and it is
signature preserving. Since our fiducial frame has signature $(++++)$ all the
frames defined with reference to it will preserve that signature.
\citet{Hestenes86:2} discusses curvature in geometric algebra formulation but
he uses a Minkowski fiducial frame with signature $(+---)$.
\section{Moving frames in 4DO}
In this section we will be looking at the frame applicable to a moving observer
and making a parallel to Lorentz transformations in special relativity. This is
a reformulation of the presentation in \cite{Almeida02:2} using the formalism
of geometric algebra. We must recall that 4DO is characterised by time being
evaluated as geodesic arc length in 4-dimensional space
\begin{equation}
    \label{eq:dt2}
    c^2 \mathrm{d}t^2 = g_{\mu \nu} \mathrm{d}x^\mu \mathrm{d}x^\nu.
\end{equation}

From this point onwards we avoid all problems of dimensional homogeneity by
using normalising factors listed in Table \ref{t:factors} for all units,
defined with recourse to the fundamental constants: Planck constant divided by
$2 \pi$ $(\hbar)$, gravitational constant $(G)$, speed of light $(c)$ and
proton charge $(e)$.
\begin{table}[bt]
\caption{\label{t:factors}Normalising factors for non-dimensional units used in
the text}
\begin{center}
\begin{tabular}{c|c|c|c}
Length & Time & Mass & Charge \\
\hline & & & \\

$\displaystyle \sqrt{\frac{G \hbar}{c^3}} $ & $\displaystyle \sqrt{\frac{G
\hbar}{c^5}} $  & $\displaystyle \sqrt{\frac{ \hbar c }{G}} $  & $e$
\end{tabular}
\end{center}
\end{table}
This normalisation defines a system of \emph{non-dimensional units} with
important consequences, namely: 1) All the fundamental constants, $\hbar$, $G$,
$c$, $e$, become unity; 2) a particle's Compton frequency, defined by $\nu =
mc^2/\hbar$, becomes equal to the particle's mass; 3) the frequent term
${GM}/({c^2 r})$ is simplified to ${M}/{r}$.

In non-dimensional units equation (\ref{eq:dt2}) above can be obtained from the
equivalent vector definition \cite{Almeida04:1}
\begin{equation}
    \label{eq:dsgen}
    \mathrm{d}s = g_\mu \mathrm{d}x^\mu,
\end{equation}
where $\mathrm{d}s$ is the displacement vector and $\mathrm{d}t^2 = \mathrm{d}s
\dprod \mathrm{d}s$. Consequently the velocity vector is defined by
\begin{equation}
    \label{eq:vgen}
    v = g_\mu \dot{x}^\mu,
\end{equation}
where "dot" over a variable means time derivative.

\begin{figure}[bt]
\centerline{\psfig{file=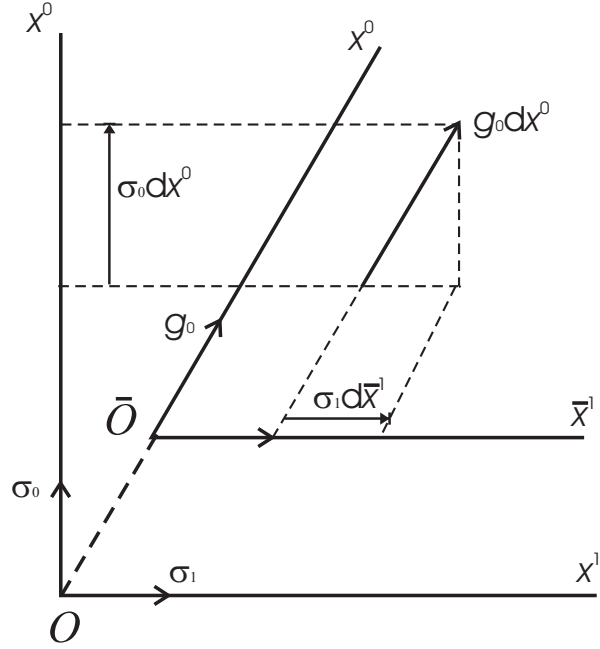, width=0.5\columnwidth}}
    \caption{\label{fig:skewframe} In his own
    frame observer $\bar{O}$ is fixed and $g_0$
    must be aligned with the velocity;
    time intervals must be the same in both frames;
    $\mathrm{d}x^0$ and time intervals must be preserved when the coordinates
    are transformed;
    the displacement
    $g_0 \mathrm{d}x^0$ can be decomposed into $\sigma_0
    \mathrm{d}x^0 + \sigma_1 \mathrm{d}\bar{x}^1$.}
\end{figure}

As an introductory example to moving observer frames we will suppose an
observer $\bar{O}$ with velocity $v = \cos \theta \sigma_0 + \sin \theta
\sigma_1$, Fig.\ (\ref{fig:skewframe}). The moving observer is obviously
stationary in his own frame and so his $\bar{x}^0$ axis must be aligned with
$v$; hence the corresponding frame vector must be obtained from $v$ by product
with a scalar: $g_0 = \lambda v$. In 4DO light is characterised by geodesics
with $\mathrm{d}x^0=0$, a necessary condition for the 3-space velocity to have
unitary norm. These displacements must be evaluated similarly by the fixed and
moving observers so their three frame vectors corresponding to 3-space must
remain unaltered: $g_i = \sigma_i$.

Any coordinate change cannot alter a displacement so that the latter can be
equally expressed in either coordinate system
\begin{equation}
    \mathrm{d}s = \sigma_\mu \mathrm{d}x^\mu = g_\mu \mathrm{d}
    \bar{x}^\mu.
\end{equation}
Applying to the case under study
\begin{equation}
    \label{eq:ds}
    \mathrm{d}s = \sigma_0 \mathrm{d}x^0 + \sigma_1
    \mathrm{d}x^1 = \lambda(\cos \theta \sigma_0 + \sin \theta \sigma_1)
    \mathrm{d}\bar{x}^0.
\end{equation}

As last argument special relativity imposes the invariance of
$(\mathrm{d}x^0)^2=\mathrm{d}t^2-\sum (\mathrm{d}x^i)^2$; in 4DO we have, from
Eq.\ (\ref{eq:dt2}), $\mathrm{d}t^2 = (\mathrm{d}x^0)^2 + \sum
(\mathrm{d}x^i)^2$. Consequently, for compatibility with special relativity, we
must require also that \footnote{In \citet{Almeida02:2} a different argument is
used to justify this invariance.}
\begin{equation}
    \label{eq:dx0}
    \mathrm{d}\bar{x}^0 = \mathrm{d}x^0.
\end{equation}
Combining Eqs.\ (\ref{eq:ds}) and (\ref{eq:dx0}) it must be $\lambda = \sec
\theta$ and
\begin{equation}
    g_0 = \sigma_0 + \tan \theta \sigma_1
    = \sigma_0 (1 + \tan \theta \sigma_{01}).
\end{equation}
The metric tensor components can be evaluated by Eq.\ (\ref{eq:metrictensor})
\begin{equation}
    g_{00} = 1 + \tan^2 \theta,~~~~
    g_{01} = g_{10} = \tan
    \theta.
\end{equation}
The evaluation of $\mathrm{d}t^2$ is unaltered because this is still the fixed
observer's time measurement performed in the moving observer's coordinates; we
call this a \emph{coordinate change}, which is different from a \emph{metric
change}. Using Eq.\ (\ref{eq:dsgen}) a general displacement is evaluated in the
two frames as
\begin{equation}
    \label{eq:dsmov}
    \mathrm{d}s = \sigma_0 \mathrm{d}x^0 + \sigma_1 \mathrm{d}x^1
    = (\sigma_0 + \tan \theta \sigma_1)\mathrm{d}x^0 + \sigma_1
    \mathrm{d}\bar{x}^1;
\end{equation}
and the coordinate conversion is immediate
\begin{equation}
    \mathrm{d}\bar{x}^1 = \mathrm{d}x^1 - \tan \theta
    \mathrm{d}x^0.
\end{equation}

The coordinate transformation for a moving observer seems incompatible with the
corresponding transformation in special relativity, which is described by a
Lorentz transformation and produces time dilation. This incompatibility is only
apparent since the moving observer has no reason to choose a \emph{skew frame}
and evaluates displacements with a standard orthonormed frame
\begin{equation}
    \mathrm{d}\bar{s} = \sigma_0 \mathrm{d}x^0 + \sigma_1
    (\mathrm{d}x^1 - \tan \theta
    \mathrm{d}x^0) = \mathrm{d}s - \sigma_1 \tan \theta
    \mathrm{d}x^0.
\end{equation}
Notice that the special relativity invariant is preserved in the moving
observer's evaluation of time
\begin{equation}
    (\mathrm{d}s)^2_{SR} = (\mathrm{d}t)^2 - (\mathrm{d}x^1)^2
    = (\mathrm{d}\bar{t})^2 - (\mathrm{d}\bar{x}^1)^2
    = (\mathrm{d}x^0)^2.
\end{equation}

The previous example restrained the analysis to movement along the $\sigma_1$
direction, described by a velocity vector with $\sigma_0$ and $\sigma_1$
components. In general a coordinate change defines a \emph{skew frame} whose
vectors result from the transformation applied to the element $\sigma_0$
\begin{equation}
    g_0 = \sigma_0 \left(1 + \frac{\sigma_0 \mathbf{v}}{v^0} \right)
    = \sigma_0 \frac{\sigma_0 v}{\sigma_0 \dprod v}\,
    = \frac{v}{\sigma_0 \dprod v} .
\end{equation}
where the bold $\mathbf{v}$ represents the 3 spatial velocity components or the
velocity vector in the non-relativistic sense. The coordinate conversion
preserves $\mathrm{d}x^0$ and for the remaining coordinates we have
\begin{equation}
    \mathrm{d}\bar{x}^k = \mathrm{d}x^k - \frac{v^k}{v^0}\, \mathrm{d}x^0.
\end{equation}
The interval evaluated by the moving observer in his orthonormed frame is
\begin{equation}
    \mathrm{d}\bar{s} = \mathrm{d}\bar{x}^\mu \sigma_\mu
    = \mathrm{d}s - \frac{\mathbf{v}}{v^0}.
\end{equation}

In conclusion, the 4DO counterpart of a Lorentz transformation in special
relativity is a two-step process.
\begin{itemize}
\item A transformation of the spatial coordinates $x^i$,
preserving the fixed observer's time measurement; the corresponding frame is
called the \emph{skew frame} and the transformation is called a
\emph{coordinate change}.

\item A \emph{jump} into the orthonormed \emph{moving frame} with the
consequent time dilation. This is mathematically a \emph{metric change} because
the length of displacements is not preserved.
\end{itemize}

A concrete situation where a moving frame had to be considered appeared in
\citet[Eq.\ (14)]{Almeida04:1}, an equation with the same form of Eq.\
(\ref{eq:dsmov}), equivalent to a frame vector
\begin{equation}
    \label{eq:hyperframe}
    g_0 = \sigma_0 + \frac{x^k \sigma_k}{x^0}.
\end{equation}
In this case the movement has a geometric cause and results from the natural
(geometric) expansion of the Universe when the hypersphere model is assumed.
\section{Electrodynamics as space curvature}
Relativistic dynamics, for cases of isotropic media, is modeled in 4DO by the
frame
\begin{equation}
    \label{eq:genframe}
    g_0 = n_0 \sigma_0,~~~~ g_j = n_r \sigma_j,
\end{equation}
where $n_0$ and $n_r$ are scalar functions of the coordinates called refractive
indices. Application of Eq.\ (\ref{eq:dt2}) with $c=1$ leads to the time
interval
\begin{equation}
    \mathrm{d}t^2 = (n_0 \mathrm{d}x^0)^2 + (n_r)^2 \sum (\mathrm{d}x^j)^2.
\end{equation}
If the refractive indices are not functions of $x^0$ the geodesics of the space
so defined can be mapped to the geodesics of the relativistic space defined by
the metric
\begin{equation}
    \mathrm{d}s^2 = \left( \frac{\mathrm{d}t}{n_0} \right)^2 - \left(
    \frac{n_r}{n_0} \right)^2 \sum (\mathrm{d}x^j)^2;
\end{equation}
as is fully demonstrated in \citet[Sec.\ 4]{Almeida04:1}. The equivalence
between 4DO and GR spaces stops when the metric is non-static, meaning that the
refractive indices are functions of $x^0$ in 4DO or $t$ in GR. The two spaces
are not equivalent either, for any displacements which involve parallel
transport.

The first question we would like to answer in this section is whether a frame
of the type defined by Eqs.\ (\ref{eq:genframe}) is adequate to model the
dynamics of a charged particle under an electric field; for this we will use
the refractive indices
\begin{equation}
    \label{eq:coulombframe}
    n_0 = 1 + \frac{q V}{m},~~~~ n_r =1,
\end{equation}
where $q$ is the charge of the moving particle, $m$ is its mass and $V$ is the
electric potential, including the fine structure constant $\alpha$. So, for
instance, the electric field of a stationary particle with charge $Q$ is
\begin{equation}
    V = \frac{\alpha Q}{r}.
\end{equation}
Notice that the refractive indices are defined for an interaction between two
charges, the same happening with the space metric. The space that we are
defining exists only for the interaction under study and it is not a
pre-existing arena where the dynamics is played.

The frame of Eqs.\ (\ref{eq:coulombframe}) produces the time definition
\begin{equation}
     \mathrm{d}t^2 = \left(1 + \frac{q V}{m} \right)^2(\mathrm{d}x^0)^2 + \sum
    (\mathrm{d}x^j)^2.
\end{equation}
Following the procedure in \citet{Almeida04:1} we will find the geodesic
equations by first dividing both members by $\mathrm{d}t^2$ and defining a
constant Lagrangian $L=1/2$
\begin{equation}
     1 = 2L = \left(1 + \frac{q V}{m} \right)^2(\dot{x}^0)^2 + \sum
    (\dot{x}^j)^2.
\end{equation}
Noting that the Lagrangian is independent from $x^0$, there must be a conserved
quantity
\begin{equation}
    \label{eq:coulombconserv}
    \left(1 + \frac{q V}{m} \right)^2 \dot{x}^0  = \frac{1}{\gamma}\, .
\end{equation}
The remaining Euler-Lagrange equations for the geodesics are
\begin{equation}
    \ddot{x}^j = \frac{q}{m}\left(1 + \frac{q V}{m} \right)\partial_j V
    (\dot{x}^0)^2;
\end{equation}
Replacing with ${\dot{x}^0}$ from Eq.\ (\ref{eq:coulombconserv})
\begin{equation}
    \ddot{x}^j =
    \frac{q  }{m \gamma^2}\left(1 + \frac{q V}{m} \right)^{-3}
    \partial_j V.
\end{equation}
In the limit of speeds much smaller than the speed of light $\gamma \rightarrow
1$. As long as $q V/m \ll 1$ the equation represents the classical dynamics of
a particle with mass $m$ and charge $q$ under the electric potential $V$.
Electric potentials typically decrease with $1/r$ and thus the parenthesis can
be taken as unity for large distances from electric field sources; for the
interaction between two electrons, considering the non-dimensional units'
normalising factors, this condition means distances considerably larger than
$2.8 \times 10^{-15}~\mathrm{m}$.

Having shown that dynamics under an electric field can be modeled by a suitably
chosen frame, we need to investigate if the same applies when a magnetic field
is present. Magnetic fields are originated by moving charges and we have seen
how a moving frame can be obtained from a stationary one by a transformation
applied to its zeroth vector. Since electric field dynamics only implies the
consideration of refractive index $n_0$, the other refractive index $n_r$
remaining unity, it is natural to admit that an electromagnetic interaction
could be modeled by the following frame
\begin{equation}
    \label{eq:emframe}
    g_0 = \sigma_0+\frac{q A^\mu \sigma_\mu}{m}\, ,~~~~ g_j = \sigma_j,
\end{equation}
where $A=A^\mu \sigma_\mu$ represents the vector potential; note that the
vector potential is referred to the orthonormed frame $\sigma_\mu$ and not to
the skew frame $g_\mu$. The equation above is equivalent to the definition of a
fiducial tensor but the vector potential approach is more convenient for the
derivations that follow.

The metric tensor elements are obtained, as usual, by Eq.\
(\ref{eq:metrictensor})
\begin{equation}
    g_{00} = \left(1+\frac{q A^0}{m} \right)^2,~~~~
    g_{0 j} = g_{j 0}= \frac{q A^j}{m}\, ,~~~~ g_{jk}= \delta_{jk}.
\end{equation}
Instead of using the geodesic Lagrangian to find its Euler-Lagrange equations,
we will follow a different procedure, starting with the velocity vector from
Eq.\ (\ref{eq:vgen}) with the frame vectors from Eqs.\ (\ref{eq:emframe})
\begin{equation}
\label{eq:vem}
    v = \frac{q A}{m}\, \dot{x}^0 + \sigma_\mu
    \dot{x}^\mu.
\end{equation}

We will now define the vector derivative
\begin{equation}
    \label{eq:nabla}
    \nabla = \sigma_\mu \partial_\mu.
\end{equation}
This is used for the derivation of an identity relative to acceleration
\begin{eqnarray}
    \dot{v} &=& \dot{v}^\mu \sigma_\mu \nonumber \\
    & = & \partial_\nu v^\mu \dot{x}^\nu \sigma_\mu \nonumber \\
    & = & \partial_\nu v^\mu v^\nu [\sigma_\nu \dprod
    (\sigma_\nu \wprod \sigma_\mu)] \\
    & = & v \dprod (\nabla \wprod v). \nonumber
\end{eqnarray}
Applying to Eq.\ (\ref{eq:vem})
\begin{equation}
    \dot{v} = v \dprod \left[ \nabla \wprod \left(\frac{q A}{m}\,
    \dot{x}^0 + \sigma_\mu  \dot{x}^\mu \right) \right].
\end{equation}
The exterior product with the second term inside the parenthesis is necessarily
null because it implies deriving the coordinates with respect to other
coordinates and not to themselves; so we have finally
\begin{equation}
    \dot{v} = \frac{q \dot{x}^0}{m}\, v \dprod (\nabla \wprod A).
\end{equation}
It is now convenient to define the Faraday bivector $F = \nabla \wprod A$. $F$
is necessarily a bivector because it is defined as an exterior product of two
vectors; we can separate it into electric and magnetic components
\begin{equation}
    F = E^j \sigma_{j0} +  B^k i \sigma_k.
\end{equation}
$E^j$ and $B^k$ are the components of electric and magnetic field vectors,
respectively; $i = \sigma_{123}$ is a special trivector incorporating the 3
spatial frame vectors, so that $i \sigma_k$ is a spatial bivector where the
component $\sigma_k$ is not present. Using boldface letters to represent
3-vectors, the Faraday bivector can be rewritten as
\begin{equation}
    F =  \mathbf{E} \sigma_0 + i \mathbf{B}.
\end{equation}
And the acceleration becomes
\begin{equation}
    \dot{v} = \frac{\dot{x}^0 q}{m}\, v \dprod \left(\mathbf{E} \sigma_0
     + i \mathbf{B}\right).
\end{equation}
This form of the Lorentz acceleration is relativistic because the norm of the
velocity vector will be kept equal to unity at all times.
\section{Maxwell's equations}
Using the vector derivative defined in Eq.\ (\ref{eq:nabla}) we write the
derivative of Faraday bivector
\begin{equation}
    \nabla F = \nabla \dprod F + \nabla \wprod F,
\end{equation}
The internal product in the second member is expanded as
\begin{eqnarray}
    \nabla \dprod F &= & \partial_k E^k \sigma_0 -
    \partial_0 E^k \sigma_k
    +(\partial_3 B^2-\partial_2 B^3) \sigma_1   \\
    && + (\partial_1 B^3-\partial_3 B^1) \sigma_2
    + (\partial_2 B^1-\partial_1 B^2) \sigma_3.
\end{eqnarray}
A careful look at the second member shows that the first term is the divergence
of $\mathbf{E}$ multiplied by $\sigma_0$. The second term is $-\partial_0
\mathbf{E}$, but we must take into account that in a stationary frame
$\mathrm{d}t = \mathrm{d}x^0$, so this term can be taken as the negative of the
time derivative in such frame. The last 3 terms represent the 3-dimensional
cross product, for which we will use the symbol "$\times$". If the bold symbol
"$\boldsymbol{\nabla}$" represents the usual 3-dimensional nabla operator, the
equation can be written
\begin{equation}
    \nabla \dprod F =  \boldsymbol{\nabla}\! \cdot \mathbf{E} \sigma_0
    + \boldsymbol{\nabla}\! \boldsymbol{\times} \mathbf{B} -
    \partial_0 \mathbf{E}.
\end{equation}
Defining the current vector $J = \rho \sigma_0 + \mathbf{J}$, for a stationary
frame, the first two Maxwell's equations are expressed by
\begin{equation}
    \nabla \dprod F = J.
\end{equation}

For the remaining two equations we have to look at the exterior product $\nabla
\wprod F$
\begin{eqnarray}
    \nabla \wprod F &=
    & (\partial_3 E^2 - \partial_2 E^3)
    \sigma_{023}
    + (\partial_1 E^3 - \partial_3 E^1) \sigma_{031} \\
    && + (\partial_2 E^1
    - \partial_1 E^2) \sigma_{012}
    +i \partial_k B^k
    + \partial_0 B^k I \sigma_k \\
    &= & i (\boldsymbol{\nabla}\! \boldsymbol{\times}\! \mathbf{E} \sigma_0
    + \boldsymbol{\nabla} \dprod \mathbf{B} + \partial_0 \mathbf{B}
    \sigma_0).
\end{eqnarray}
The remaining two equations are expressed as $\nabla \wprod F = 0$ and the four
equations can then be condensed in the very compact form
\begin{equation}
    \nabla F = J.
\end{equation}
Additionally, it is possible to choose $A$ such that $\nabla \dprod A =0$, so
that an equivalent form of the four equations is
\begin{equation}
    \nabla^2 A = J,~~~~(\nabla \dprod A = 0.)
\end{equation}

Solutions of special interest are those where the second member is null $(J =
0)$. Some solutions can be found easily by splitting $\nabla^2 A$ in its 0th
and spatial components
\begin{equation}
    \nabla^2 A = \partial_{00} A + \boldsymbol{\nabla}^2 A,
\end{equation}
where the second term is a standard 3D Laplacian. One possible solution arises
immediately when $\partial_{00} A = \omega^2 A$ with $\omega$ real;
\begin{equation}
    \boldsymbol{\nabla}^2 A = -\omega^2 A,
\end{equation}
results in a well known Helmholtz equation. Looking for a particular solution
we assume only $x^3$ and $x^0$ dependence to find
\begin{equation}
    \label{eq:sol1}
    A = A_0 \mathrm{e}^{- \omega x^0 -u \omega x^3}.
\end{equation}
Where $u^2 = -1$ and $u$ must commute with $\sigma_3$. The two possibilities
are either $u=i$ or $u=i\sigma_3$; we will choose the latter. Since we know
that $A$ is a vector it must be
\begin{equation}
    A_0 =\alpha \sigma_1+ \beta \sigma_2;
\end{equation}
an eventual component aligned with $\sigma_0$ has been ignored.

In order to make interpretation of Eq.\ (\ref{eq:sol1}) easier we shall now
write it in a slightly modified form
\begin{equation}
    A = \mathrm{e}^{- \omega x^0} \mathrm{e}^{i\sigma_3 \omega x^3/2}
    (\alpha \sigma_1+ \beta \sigma_2) \mathrm{e}^{-i\sigma_3 \omega x^3/2}.
\end{equation}
It is now apparent that the first exponential factor is evanescent in the
positive $x^0$ direction and grows to infinity in the negative direction; this
is an uncomfortable situation that will be resolved below. The remaining
factors represent a vector that rotates in the $i\sigma_3$ plane, along the
$x^3$ direction, with angular frequency $\omega$. With an adequate choice for
the origin of $x^3$ it is possible to make $\beta=0$.

Before we proceed to the analysis of evanescence along $x^0$ let us consider
consider the expression for $A$ when $x^0$ is constant; setting $\alpha' =
\alpha \exp(-\omega x^0)$ it is
\begin{equation}
    A = \alpha' \sigma_1 \mathrm{e}^{-i\sigma_3 \omega x^3}.
\end{equation}
Along the direction normal to the rotation plane we can define time to be equal
to the distance traveled, $t = x^3$ and so $A$ can also be expressed as a time
function
\begin{equation}
    A = \alpha' \sigma_1 \mathrm{e}^{-i\sigma_3 \omega t}.
\end{equation}
In spacetime formulation time is a coordinate and we express the fact that $t =
x^3$ by giving $A$ opposite dependencies in the two variables
\begin{equation}
    A = \alpha' \sigma_1 \mathrm{e}^{-i\sigma_3 \omega (x^3-t)}.
\end{equation}
This is the spacetime formulation for a circularly polarised electromagnetic
wave \cite{Lasenby99}.

In \citet{Almeida04:1} the Universe is modeled as an hypersphere whose radius
is the coordinate $x^0$. For displacements that are small compared to the
hypersphere radius it is possible to use length coordinates rather than angles,
resulting in the $g_0$ frame vector of Eq.\ (\ref{eq:hyperframe}). A comparison
between this equation and Eq.\ (\ref{eq:emframe}) shows that the Universe's
expansion is the source of a vector potential $A = x^k\sigma_k /x^0$; the
derivations in the previous section show that there are solutions to Maxwell's
equations that force the vector potential to rotate and the frame vector $g_0$
with it. These solutions force $x^0$ to stay constant and so they are
constrained to hyperspherical surfaces of constant radius. This solves the
unanswered question raised in \citet{Almeida04:1} whereby photons were
constrained to great circles on the hypersphere, rather than to geodesic
straight lines.

We can now turn our attention to the evanescence problem or rather to the
problem of an amplitude growing exponentially in the negative $x^0$ direction.
This is the direction pointing towards the hypersphere centre and suggests that
the approximations made in order to use flat space are the source of the
inconsistency. Recalling that in the hypersphere model the Universe is
expanding at the speed of light, we can expect that there are resonating modes
inside the hypersphere, which are evanescent to its outside due to continuity
on the hypersphere border. A correct solution of Maxwell's equations should not
ignore this fact and will be pursued in forthcoming work. The resonating modes
are so closely spaced with the current size of the Universe that we can
actually consider a continuous distribution, but they manifest themselves as
electromagnetic waves when space is artificially flattened.
\section{Conclusion}
The formalism of geometric algebra was used in this paper as useful
mathematical tool for writing complex equations in a compact form and
facilitate their geometrical interpretation. Although most readers will not be
familiar with this algebra, we think that the simplifications achieved through
its use are well worth the extra effort of learning some of its rudiments. A
reformulation of Maxwell's equations in 4-dimensional Euclidean space is the
goal of this work but an introduction to the discipline of 4-dimensional optics
was felt necessary, namely to explain the new transformation for moving frames
and its relation to Lorentz transformations in special relativity.

Through the consideration of an electromagnetic vector potential embedded in
one of the frame vectors, the paper shows that relativistic electrodynamics can
be derived from curved space geodesics. In this way the Lorentz force acquires
the characteristic of an inertial force, which is best described by inertial
movement in curved space. Maxwell's equations are then established in this
space, with geometric algebra achieving the feat of their condensation in the
equation $\nabla^2 A = J$. Solving the equations for free space leads to
rotation of frame vector $g_0$ on a plane lying on 3-space, with rotation
progressing at arbitrary frequency along the direction normal to the rotation
plane. These solutions are evanescent on the positive $x^0$ direction and grow
to infinity in the opposite direction. The inconsistency is attributed to
artificial flattening of the hypersphere space proposed in previous work
\cite{Almeida04:1} and further work is suggested to fully clarify this point.
  \bibliographystyle{unsrtbda} 
  \bibliography{Abrev,aberrations,assistentes}   
\end{document}